\documentclass[a4paper,11pt]{article}
\pdfoutput=1 % if your are submitting a pdflatex (i.e. if you have
\usepackage{jinstpub} % for details on the use of the package, please
\title{Discriminating cosmic muons and radioactivity using a liquid scintillation fiber detector}

\author[a,b,1]{Y.P. Zhang,\note{Corresponding author.}}
\author[a]{J.L. Xu,}
\author[a]{H.Q. Lu,}
\author[a]{P. Zhang,}
\author[a]{C.C. Zhang,}
\author[a]{and C.G. Yang}

\affiliation[a]{Institute of High Energy Physics, Chinese Academy of Sciences, Yuquan Road, Beijing,100049, China}
\affiliation[b]{University of Chinese Academy of Sciences, Yuquan Road, Beijing, 100049, China}

\emailAdd{ypzhang1991@ihep.ac.cn}

\abstract{In the case of underground experiments for neutrino physics or rare event searches, the background caused by cosmic muons contributes significantly and therefore must be identified and rejected. We proposed and optimized a new detector using liquid scintillator with wavelenghth-shifting fibers which can be employed as a veto detector for cosmic muons background rejection. From the prototype study, it has been found that the detector has good performances and is capable of discriminating between muons induced signals and environmental radiation background. Its muons detection efficiency is greater than 98$\%$, and on average, 58 photo-electrons (p.e.) are collected when a muon passes through the detector. To optimize the design and enhance the collection of light, the reflectivity of the coating materials has been studied in detail. A Monte Carlo simulation of the detector has been developed and compared to the performed measurements showing a good agreement between data and simulation results.}

\keywords{ Liquid detectors, Scintillators, Detector design}

\begin{document}
\maketitle
\flushbottom

\section{Introduction}
\label{sec:intro}
\begin{figure}[htbp]
  \centering
 % \begin{center}
  \includegraphics[width=8cm]{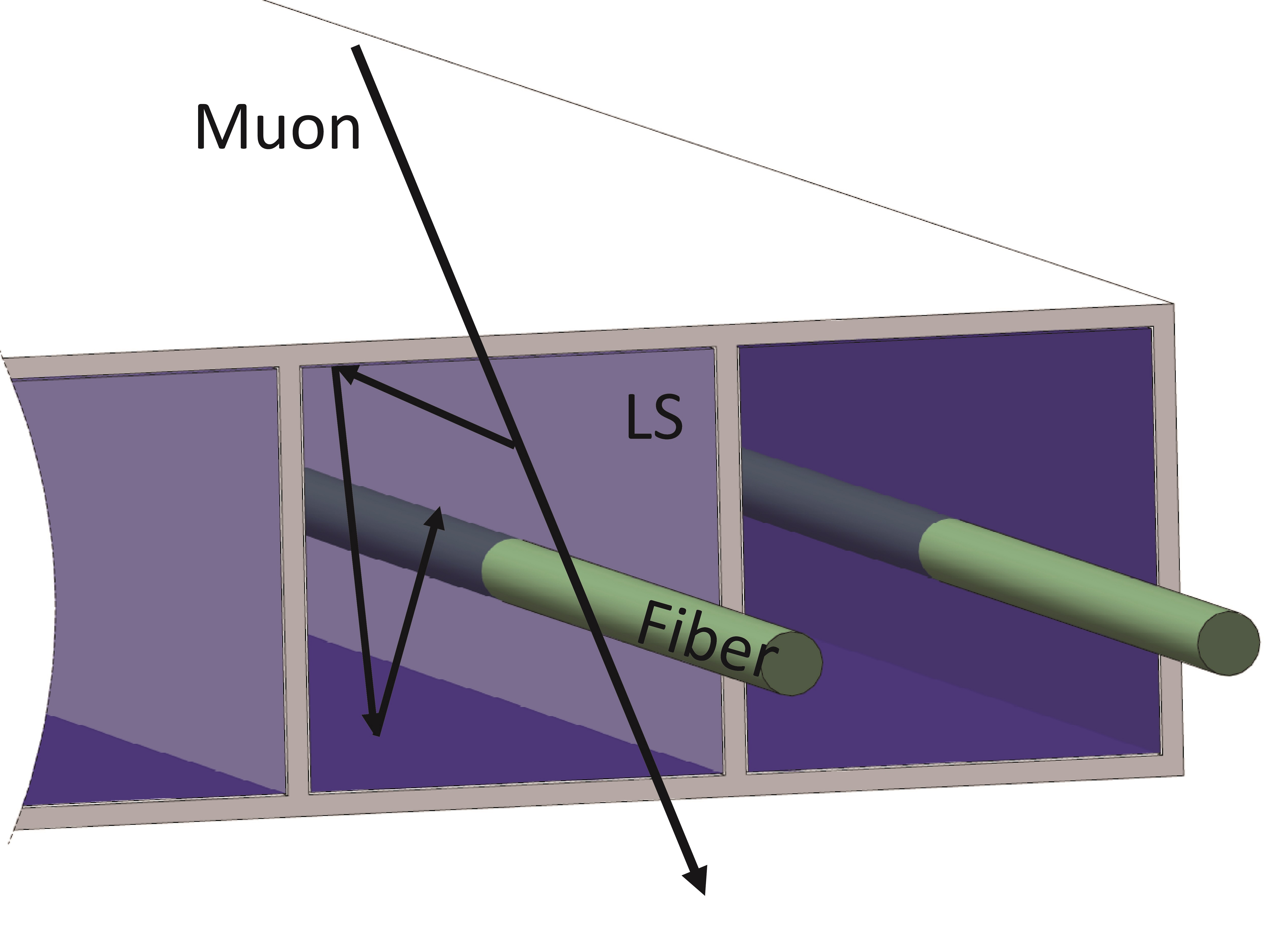}
  \caption{\label{fig:detPrinciple} LSF detector working principle.}
 % \end{center}
  \end{figure}

The basic working principle of a Liquid Scintillator with Fibers (LSF) detector is shown in Fig.\ref{fig:detPrinciple}. An optically sealed container is filled with Liquid Scintillator (LS), and a WaveLength Shifting (WLS) fiber placed in the centre of the container and connected at both ends to standard Photomultiplier Tubes (PMT). Ionizing radiation passing through the LS will generate optical photons, which will be eventually collected by the fibers and read out at both ends by the PMTs. Multiple containers are put together to form a layer. An arrangement of multiple layers makes a particle detector with tracking capabilities.

The LSF detectors have been successfully employed in neutrino physics experiments thanks to their excellent performance and moderate cost. It was one of the detector option for the MINOS~\cite{MINOS} experiment at Fermilab and it has been selected, as an example, for the NO$\nu$A~\cite{NOVAdesign,NOVAdet} long-baseline neutrino experiment. The NO$\nu$A detector is made of 344,000 cells of extruded, highly reflective plastic polyvinyl chloride (PVC) filled with LS and read out by a WLS fiber connected to Avalanche Photo-Diodes. Using the pattern of light seen by the photo-detectors, what kind of neutrino caused the interaction and what its energy was can be determined. One of the main reasons to use such detector technology for neutrino physics is due to the capability of covering large areas with moderate costs obtaining very good tracking performances. This is expecially relevant for underground neutrino or dark-matter search experiments where large areas have to be covered.

As for the underground large-scale particle physics experiments, due to low signal rates and large volume of detector, the background caused by cosmic ray muons is the main background. Although the purpose of NO$\nu$A detectors is not to measure cosmic muons, an optimized LSF detector also can be used as veto detector which can reject radioactivity background from the surrounding rock and tag cosmic muons in underground neutrino experiments or underground dark matter experiments. According to cosmic muons event rates and radioactivity background level, the thickness of the LS and number of LSF detectors can be adjusted so that the detector gains the ability to distinguish between cosmic ray signals and signals from background radioactivity. Recently, the LSF technology was proposed as an option to be used for the veto detector of the Jiangmen Underground Neutrino Observatory (JUNO) experiment~\cite{juno_concept}. The JUNO~\cite{juno} detector, under construction in a 700 meters-water-equivalent (m.w.e.) underground lab, will explore the neutrino mass hierarchy by measuring the reactor anti-neutrino spectrum in a large (20 kt) liquid scintillator detector. JUNO is a multi-purpose detector and will address several neutrino physics topics, such as measurement of the neutrino oscillation parameters, and it will explore the solar, atmospheric and supernova neutrinos physics. Background rejection (environmental radioactivity and cosmic rays induced background) is therefore a crucial issue to fully exploit the physics potential of JUNO. Neutrino detectors are located in a deep underground lab and combined with veto detectors for cosmic muons background rejection.

If the LSF detectors are compared to other tracking devices, for instance Resistive Plate Chambers (RPC)~\cite{RPC}, the former show much smaller dependence on environmental parameters (i.e. temperature and humidity). Moreover the maintenance and operation of an LSF detector is easier compared to gaseous detectors. The successful production and excellent performance of LS in Daya Bay neutrino experiment~\cite{dayabaypaper} prove that the condition is mature for LSF detector fabrication in China.

In the following paragraphs, the structure of an LSF prototype is described and its performance results are discussed in details. The results show that the technology is perfectly adequate for the requirements of a veto detector for underground physics experiments.

\section{Experimental setup}

\subsection{Prototype detector construction}

\begin{figure}[htbp]
  \centering
  %\begin{center}
  \includegraphics[width=7cm]{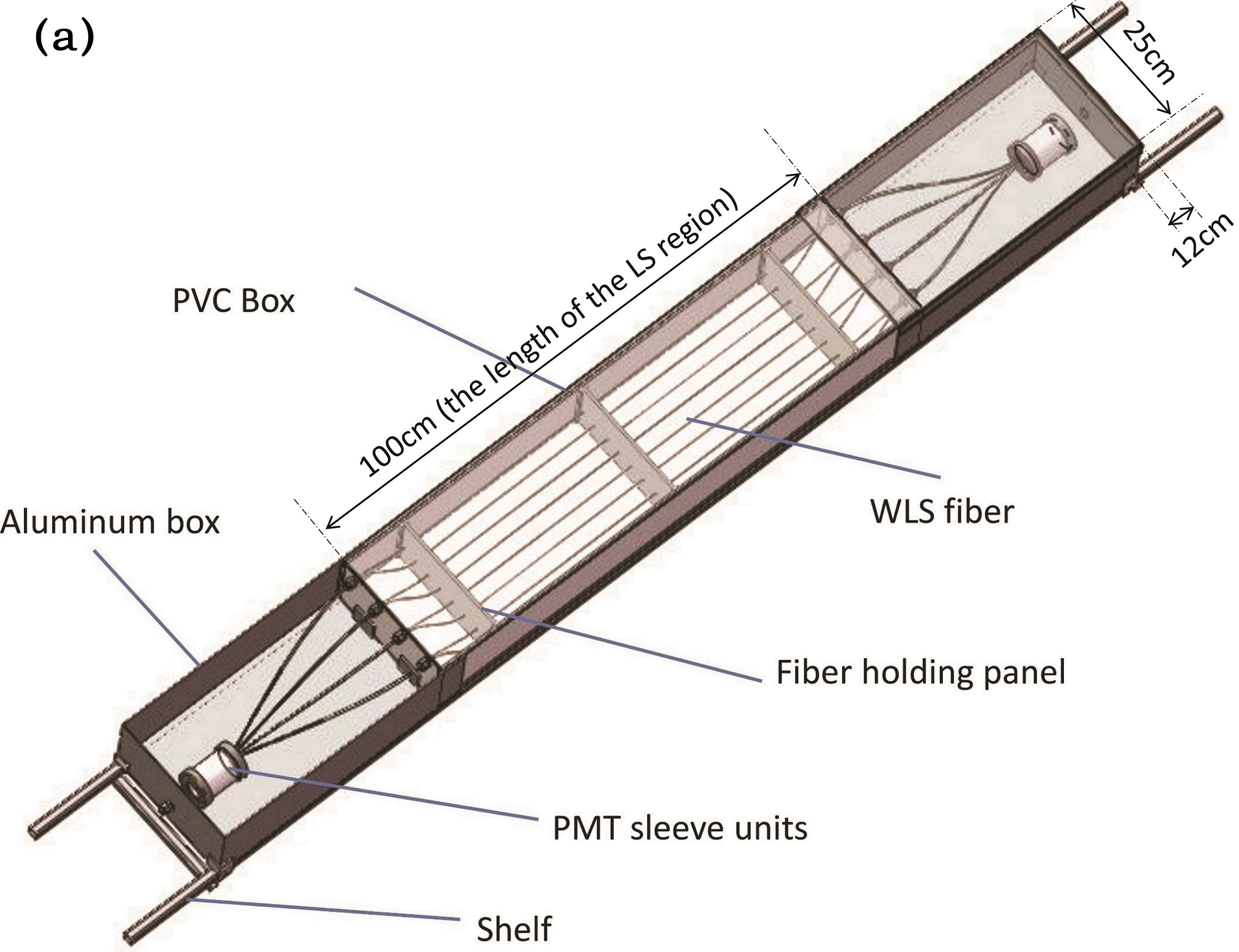}
  \qquad
  \includegraphics[width=7cm]{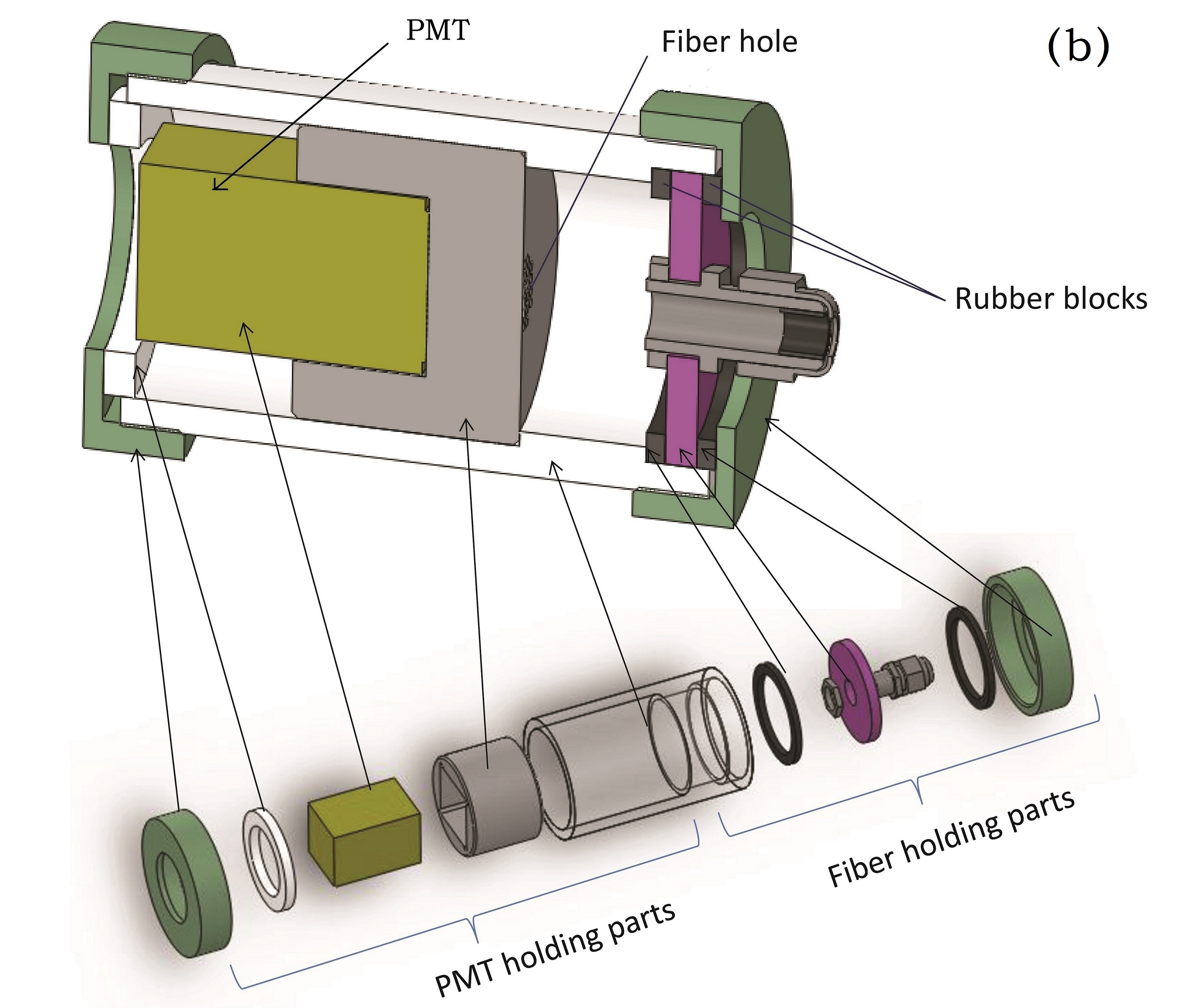}
  \caption{\label{fig:prototype} (a) LSF prototype detector schematic. (b) The design of the PMT sleeve units. }	
  %\end{center}
  \end{figure}

 Figure \ref{fig:prototype}(a) shows the schematic of an LSF prototype module. The container box, with approximate dimensions of 100 $\times$ 25 $\times$ 12 cm$^{3}$, is filled with LS~\cite{dayabayLS1, dayabayLS2, LSatt_dyy}. The BCF-92 type wavelength shifting fibers from Saint-Gobain company~\cite{fiber} with a diameter of 1.5 mm are placed inside the container. The fiber is readout by two Hamamatsu PMTs (R7600U-100-M4) with four anodes and a high gain of about 5 $\times$ 10$^{6}$~\cite{pmtR7600}. The base material for the container is PVC, used due to a high degree of compatibility with LS and its relatively cheap price. To enhance the light reflectivity of the internal container walls, the PVC has been mixed with titanium dioxide. It is well known that pure PVC and TiO$_{2}$ are compatible with LS, but if employed in a mixture with other auxiliary materials, it may reduce or even jeopardize the compatibility with LS. Therefore the ratio of these materials has been adjusted to make sure that the PVC strength is maintained and performance remains high after the application of TiO$_{2}$ and lesser amounts of other auxiliary materials.

Several PVC samples from Rifeng factory~\cite{rifeng} in Foshan city have been studied and tests of their LS compatibility and reflectivity have been performed. Eventually, PVC type 1071 was chosen as the material for the prototype container. The reflectivity of PVC type 1071 is shown in Fig. \ref{fig:reflectivities}.

As shown in Fig. \ref{fig:prototype}(a), the internal area of the LS container could be split into several narrow units by the PVC panels,  which can be used to test the position resolution of different width containers. Eight WLS fibers were uniformly immersed into the LS to collect light. According to the results of MINOS~\cite{MINOS}, variance in the fiber placement at the corners and central positions can cause a 10$\%$ variance in the p.e. collection. In order to reduce light loss, the flat end of the fibers was lapped. After fiber assembly, both ends of the PVC container were covered in an aluminum box. To guarantee a good coupling to the fibers, the PMT was first put into a sleeve and afterwards inside an aluminum box. The PMT sleeve unit was designed and custom-made. The design of the sleeves is shown in Fig. \ref{fig:prototype}(b). As shown in Fig. \ref{fig:prototype}(b), the sleeve is composed of two parts including PMT holding part and fibers holding part. First of all, the optical fibers passed through the rubber blocks and were kept in the fiber hole. Then the PMT was placed in the appropriate position of the sleeve, and the sleeve shell was screwed tightly so that there was a good coupling between the optical fiber and PMT. Two LEDs were placed close to both PMTs for gain calibration. The readout signal cables and the PMT power supply cables exit from the aluminum boxes and the whole prototype module was made light-tight. Finally, the LSF prototype module was placed between three layers of plastic scintillators which were used as a cosmic ray telescope to trigger signals from passing muons. The plastic scintillator has approximate dimensions of 93 $\times$ 15 $\times$ 5 cm$^{3}$. The signals were recorded using an oscilloscope.

Throughout the detector construction and assembly process care was taken to maintain a high level of cleanliness, to avoid impurities. All parts were cleaned using deionized water in an ultrasonic bath.

\subsection{PMT calibration}

With the PMT put into a dark box, a dark current of about 321 $\mu$A (max 400 $\mu$A) with an applied voltage of 880 V (max 900V) has been measured. The two PMTs coincidence noise rate, approximately 0.55 Hz, has been measured with a 200 ns time window at a 3 mV signal threshold; on increasing the threshold to 4 mV, the coincidence rate reduces to less than 0.01 Hz. The PMT calibration was performed in a dark box using an LED which was driven by a pulse generator. The PMT waveforms were recorded by oscilloscope at a 100 ps time resolution. For each PMT, different LED intensities were used to calibrate the PMT single p.e. peak and to get reliable results which is average value of different gains under different LED intensities.

\begin{figure}[htbp]
  \centering
  %\begin{center}
  \includegraphics[width=8cm]{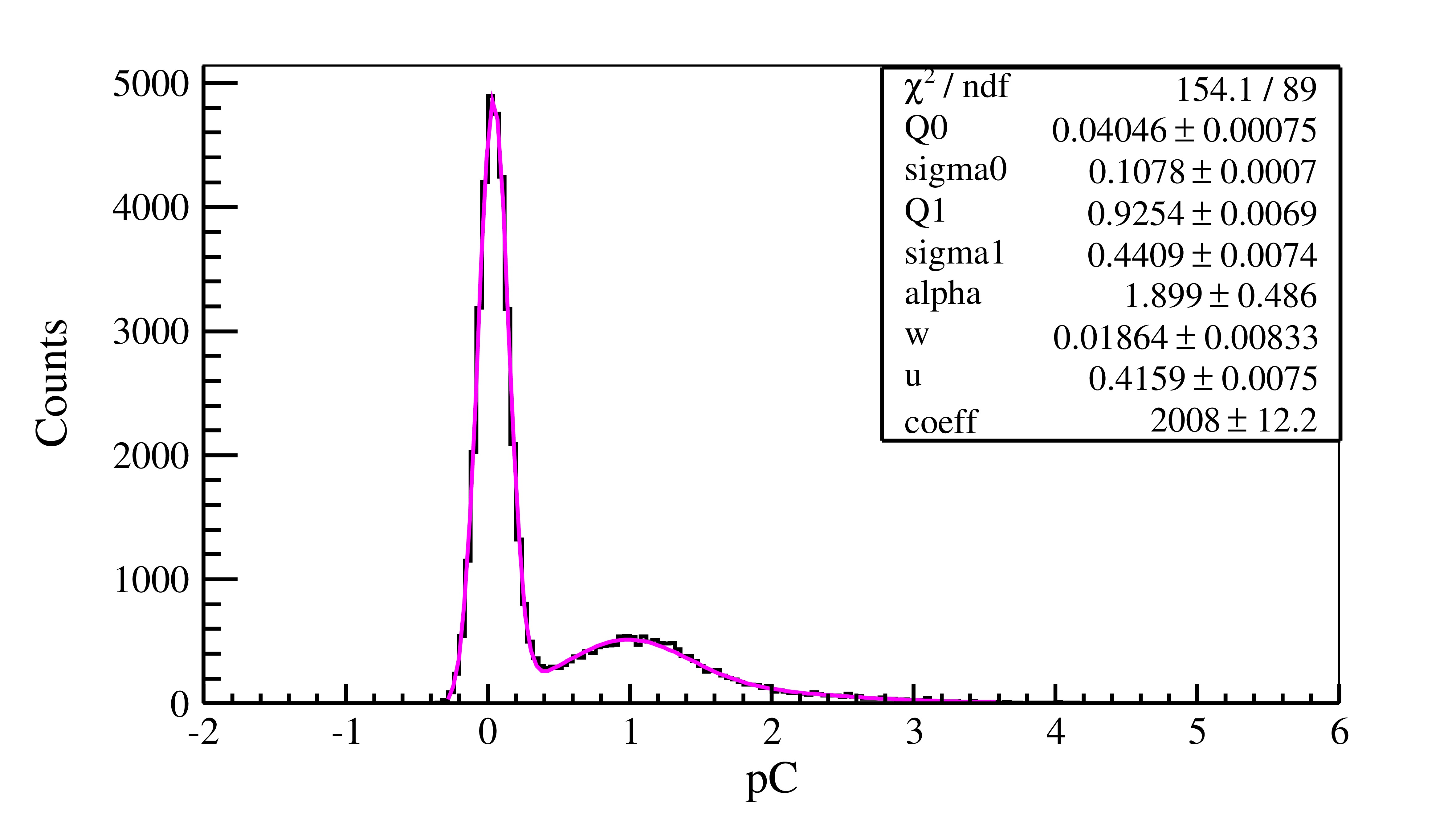}
  \caption{\label{fig:speGain} The single p.e. spectrum of the PMT was obtained by LED calibration in the LSF prototype detector. The spectrum was fitted using a PMT response function described in \cite{PMTCalib}.}
  %\end{center}
  \end{figure}

Figure \ref{fig:speGain} shows the signal PMT charge spectrum after waveform integration. Assuming a certain PMT response model~\cite{PMTCalib}, by fitting the single p.e. spectrum, the single p.e. peak with charge Q1 of about 0.93 pC is obtained. The gain calibration was performed twice, before and after each measurement. The gains of two PMTs are (4.9 $\pm$ 0.1) $\times$ 10$^{6}$ and (5.6 $\pm$ 0.3) $\times$ 10$^{6}$, respectively.

\subsection{Detector response and efficiency for muons}

Cosmic ray muons pass through the LS, deposit some energy which produces scintillation photons, and these photons are collected by the fibers and transmitted to the PMTs. If the number of collected p.e. is high enough, setting an appropriate discrimination threshold allows to obtain a better signal-to-noise ratio. Several factors influence p.e. collection, such as LS attenuation length, LS thickness, container reflectivity, WLS fiber attenuation length, PMT quantum efficiency, the nature of the couplings between WLS fibers and PMT, and others. The photons can only travel a few dozen centimeters after several reflections because of the prototype's compact geometry, and the LS attenuation length is greater than 10 m according to the previous study~\cite{LSatt_dyy}, therefore the influence caused by the LS attenuation length can be neglected. Most photons are absorbed at the walls of the LS container.

\begin{figure}[htbp]
  \centering
  %\begin{center}
  \includegraphics[width=8cm]{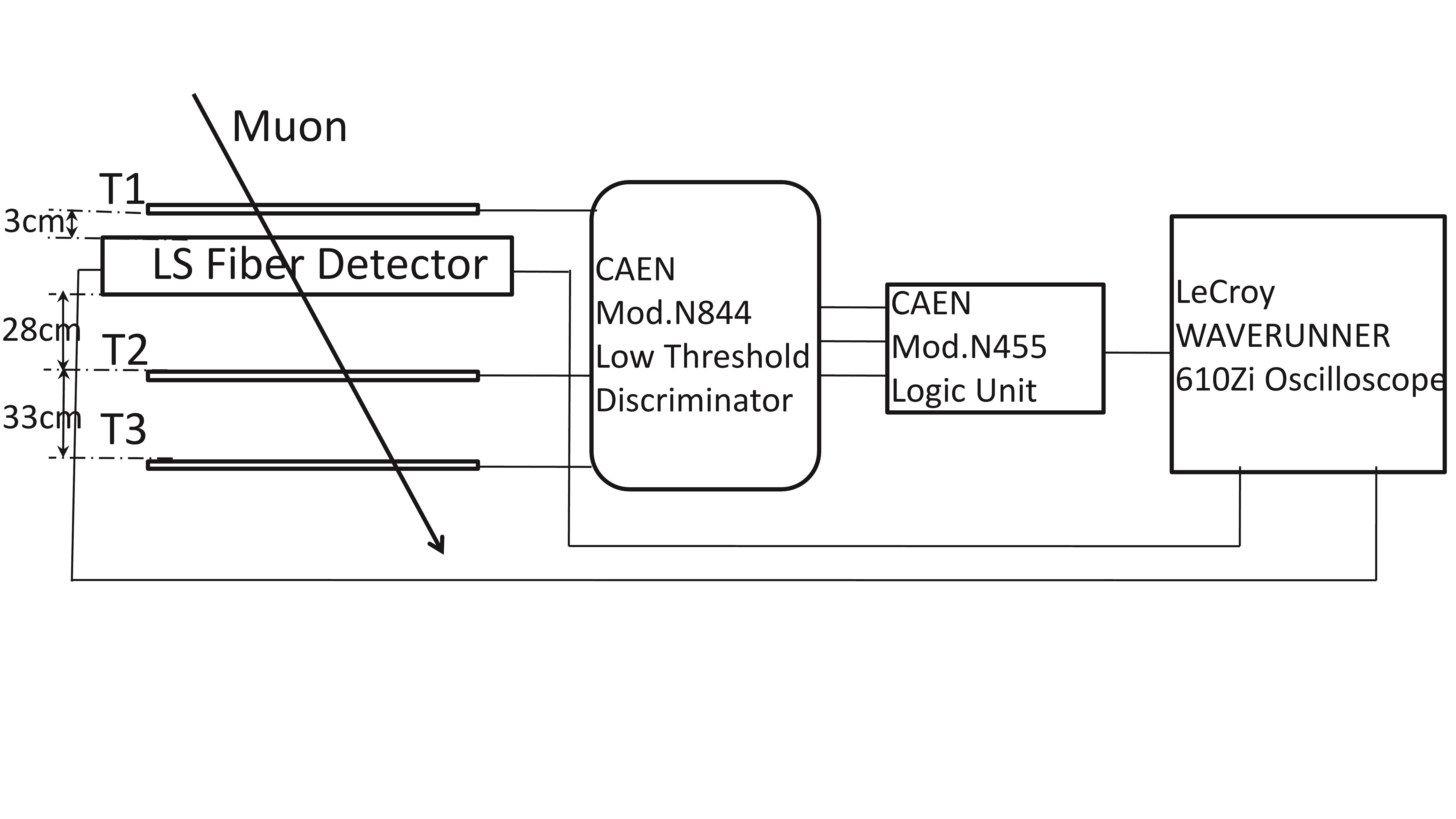}
  \caption{\label{fig:detlogic} Logical diagram for the signal waveform recording and the layout drawing with distance of the plastic scintillators. T1, T2 and T3 indicate the plastic scintillators of the cosmic ray telescope.}
  %\end{center}
  \end{figure}

The signals coming from three scintillators then pass through a low threshold discriminator module with threshold of 25 mV. The discriminator creates a coincident signal that is used as a trigger input for the oscilloscope. Figure \ref{fig:detlogic} shows the logical diagram for the waveform signal recording and the layout drawing with distance of the plastic scintillators.

\begin{figure}[htbp]
  \centering
  %\begin{center}
  \includegraphics[width=8cm]{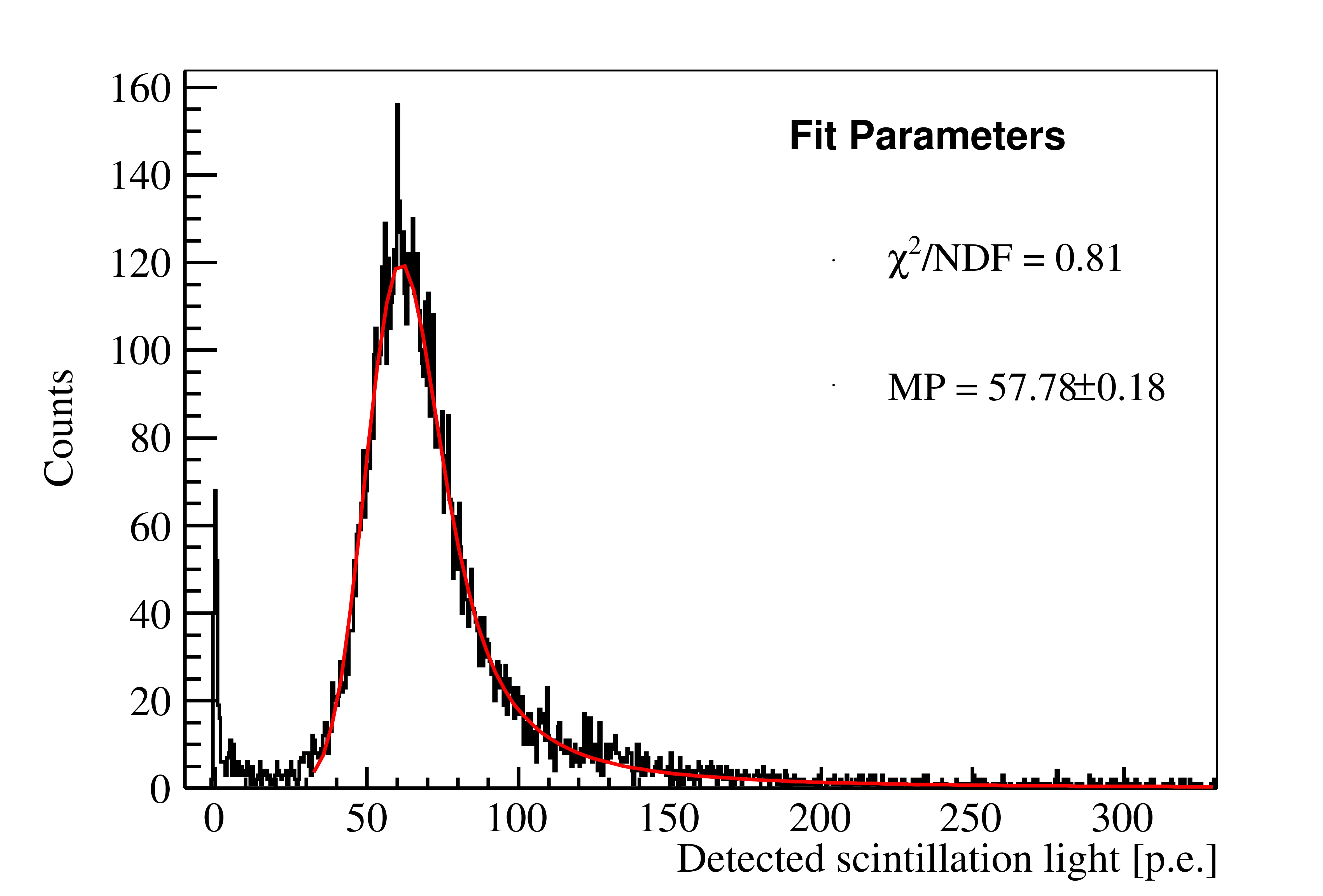}
  \caption{\label{fig:totalpeOfMu} Detector response for passing muons}
  %\end{center}
  \end{figure}

When muons pass through several centimeters of material, the ionization energy loss obeys a Landau distribution while the number of optical photons the LS emits follows the Gaussian function. As can be seen in Figure \ref{fig:totalpeOfMu}, the energy spectrum collected for the passing muons has been fitted to a Landau convoluted with a Gaussian distribution. The most probable value for the number of photo electrons, with 8 cm LS depth, is 58 p.e. This value was obtained using PVC as reflector in the LSF, without the use of additional reflector materials.

As shown in Figure \ref{fig:detlogic}, one plastic scintillator pad was placed above the prototype module and two were placed underneath. All telescopes were vertically aligned along the shelf. The sensitive areas of telescope pads are smaller than the LS' active area, and keep all muons selected to pass through the LS. The detection efficiency and its statistical error have been calculated using the following function:

\begin{equation}
\label{equ:eff}
\epsilon = \frac{N_{T1\&T2\&T3\&PMT1\&PMT2}}{N_{T1\&T2\&T3}}
\end{equation}

\begin{equation}
\label{equ:sigma}
\sigma = \sqrt{\frac{\epsilon(1-\epsilon)}{N_{T1\&T2\&T3}}}
\end{equation}

In the above function, T1, T2 and T3 are three telescope pads. PMT1 and PMT2 are the PMTs at the both ends of the LSF detector. The numerator in equation \ref{equ:eff} is the counts of the coincident signal among the three telescope pads and the two PMTs over a 100 ns time window. The denominator is the counts of coincident signal among the three telescopes pads. The signals coming from the telescope pads and the PMTs are transmitted through a low threshold discriminator with 25 mV thresholds for the telescope pads and 4 mV for the LSF detector. A higher threshold was used for the trigger to reduce coincidence noise from the plastic scintillator. The measured detection efficiency is about 98.6 $\pm$ 0.1$\%$. The small degree of inefficiency is due to the dead areas in the LSF detector: the fiber holding panel structure in the LS and the bends of the WLS fibers before they reach out of the LS.

\section{Light yield of the LSF detector }

\subsection{Influence of liner reflectivity}

\begin{figure}[htbp]
  \centering
  %\begin{center}
  \includegraphics[width=8cm]{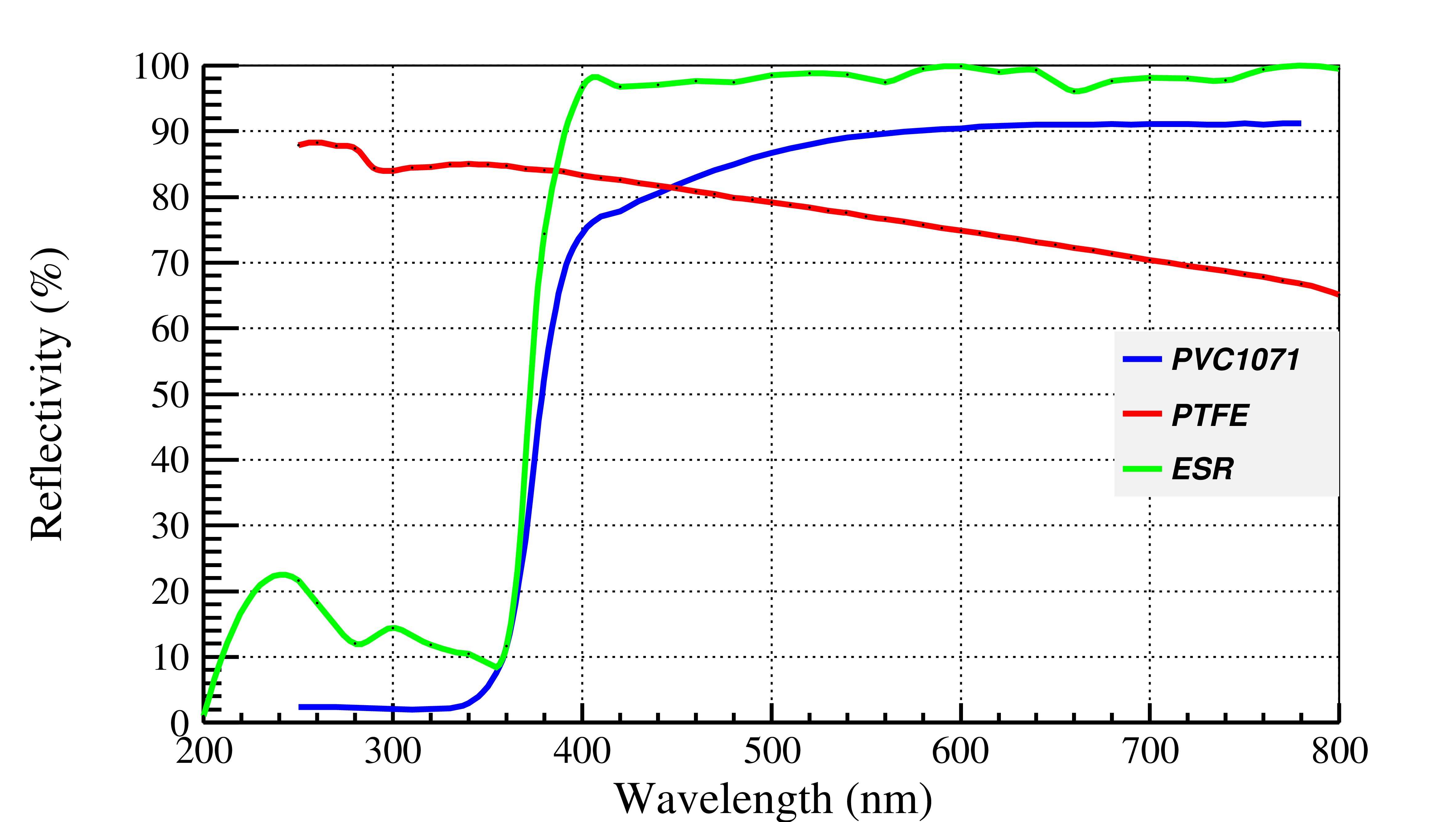}
  \caption{\label{fig:reflectivities} Different material reflectivity. The reflectivity of a special type of PVC, type 1071 produced by Rifeng Co. Ltd \cite{rifeng}, and PTFE are measured in the air by International Institute of Metrology, China \cite{nim}. ESR reflectivity data come directly from the producer's data sheet.}
  %\end{center}
  \end{figure}

In order to increase the collected number of p.e., different materials compatible with the LS were added as reflectors in the LSF container. Enhanced specular reflector (ESR) and polytetrafluoroethylene (PTFE) were selected as coating materials for the inner walls of the container. The p.e. yield of different materials could be measured by fitting the energy spectrum of detector response when muons pass through the LS. The p.e. collection results of different materials are shown in table \ref{tab1:pecollections}. Figure \ref{fig:reflectivities} shows the reflectivity of PVC, ESR and PTFE. The PTFE reflectivity decreases with increasing wavelength, while the ESR and PVC reflectivity show a step-like function with very low values below 380nm, and reaching a high value with increasing wavelength.

\begin{figure}[htbp]
  \centering
  %\begin{center}
  \includegraphics[width=8cm]{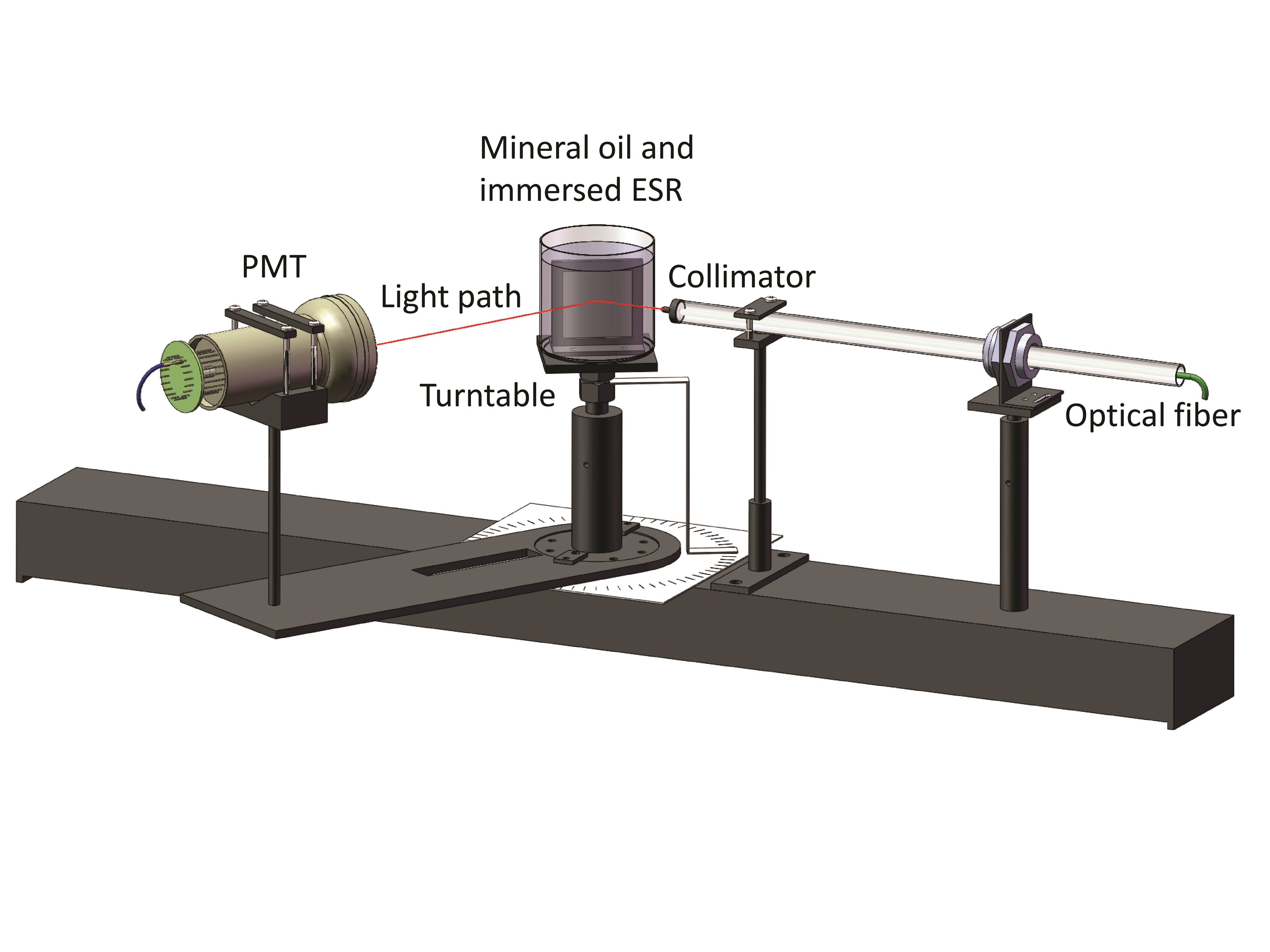}
  \caption{\label{fig:ESRdet} ESR reflectivity measurement schematics.}
  %\end{center}
  \end{figure}

The wavelength peak of the Daya Bay LS emission spectrum lies between 400 and 440 nm, but the p.e. collection of ESR with the highest measured reflectivity turned out to be much lower than that of PVC and PTFE. Previous studies~\cite{ESRreflect} have shown that ESR reflectivity depended on the photon incidence angle when ESR was immersed in mineral oil. Therefore an experiment has been performed to measure ESR film reflectivity. The schematic of ESR reflectivity measurement is shown in Fig. \ref{fig:ESRdet}: a photon beam produced by an LED emitting at an appropriately chosen wavelength was transmitted through an optical fiber, focused and made parallel after exiting the optical fiber. Photons reflected by the ESR are then detected by the PMT.

\begin{figure}[htbp]
  \centering
  %\begin{center}
  \includegraphics[width=8cm]{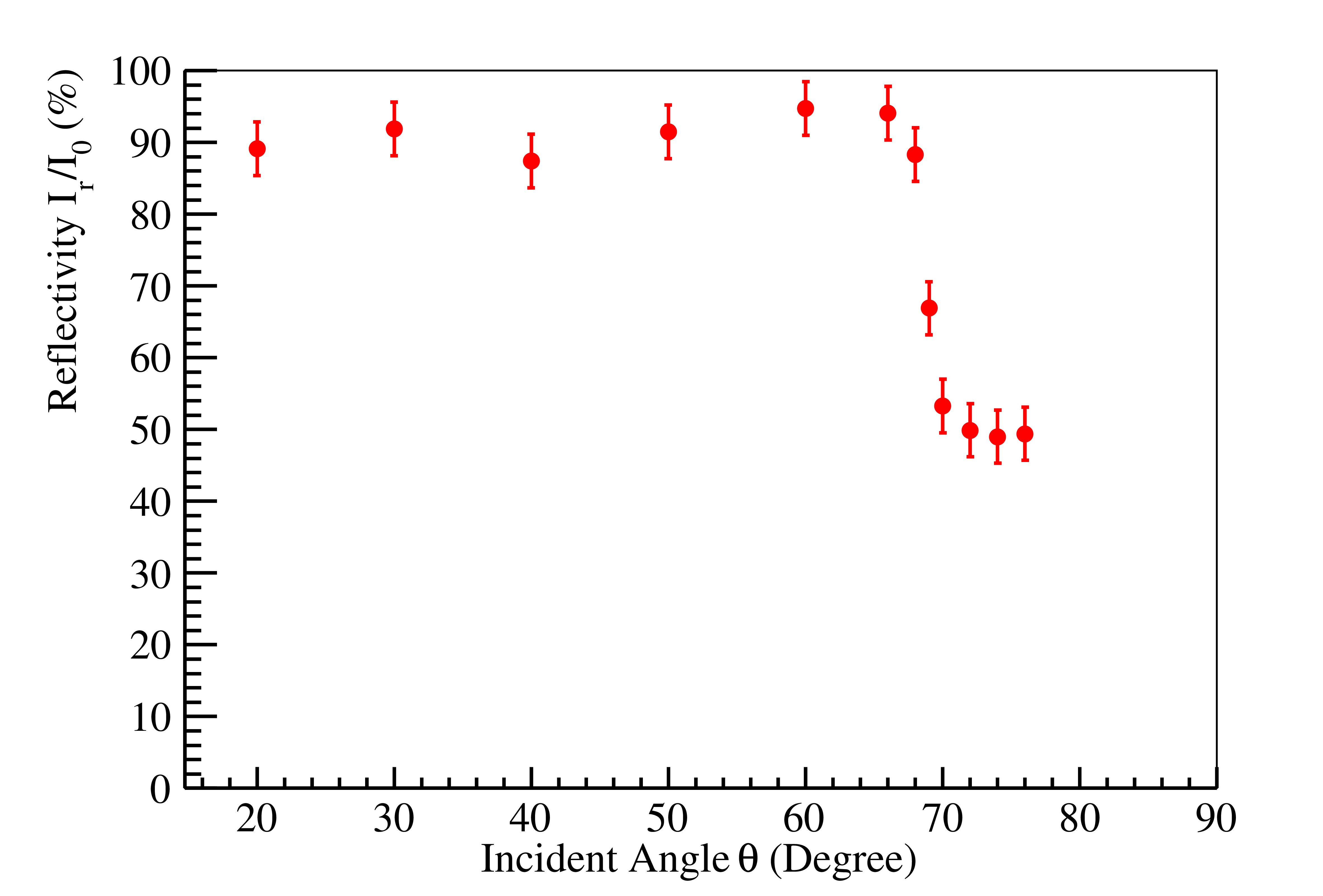}
  \caption{\label{fig:420nmRef} ESR reflectivity with different incidence angles. The optical source is an LED with a center value of 416nm (sigma 8nm). For every point in the diagram, the error on the angle is about 0.5 degrees.}
  %\end{center}
  \end{figure}

During this process, a reference measurement has been taken before immersing the ESR sample in the LS. After vertically inserting the ESR sample in a quarz glass cup filled with mineral oil, the reflectivity measurement is taken. The PMT can be rotated at any given angle to match and measure the reflected light. The collected charge is then integrated through a QDC (Charge-to-Digital Converter). The measured reflectivity, R, can be expressed by the following equation:

\begin{equation}
\label{equ:reflectivity}
R = \frac{Q_{2}-Q_{0}}{Q_{1}-Q_{0}}
\end{equation}

where $Q_{0}$ is the measured charge with the LED off, and thus is the background normalization of the measurement. $Q_{1}$ is the charge without ESR which expresses incident light intensity, and $Q_{2}$ is the reflected light charge. Figure \ref{fig:420nmRef} shows the measured ESR reflectivity as a function of the incident photon's angle.

ESR reflectivity reaches about 90$\%$ when the incident angle becomes smaller than 65 degrees and exhibits a sharp decrease to about 50$\%$ for larger incident angles. The reflectivity measurement was repeated systematically for other wavelength LEDs, ranging from 365nm to 600nm at approximate 20nm intervals. A similar behavior was observed for each LED as a function of the incident angle. As a final comment on reflectivity, when the incident angle becomes smaller than 65 degrees, Figure \ref{fig:420nmRef} shows that the obtained value is lower than that given in the ESR data sheet. The discrepancy is likely caused by light transmission losses and diffusion reflections when the light goes through different media.

In order to better understand the performance of the detector, a Monte Carlo (MC) simulation was implemented using the Geant4~\cite{Geat4}. The measured reflectivity data were input into simulation. A simulation of PTFE, PVC and ESR coating of the internal container walls was implemented and created simulation results which were consistent with the measurements. Simulation results are shown in table \ref{tab1:pecollections}.

\begin{table}[htbp]
\centering
%\begin{center}
\caption{ \label{tab1:pecollections} P.e. collection results of different materials obtained by measurement and simulation.}
%\footnotesize
\smallskip
\begin{tabular*}{80mm}{c@{\extracolsep{\fill}}ccccc}
%\begin{tabular}{|lr|c|}
\hline
%\toprule
Material &  Exp. (p.e.) & Sim. (p.e.)\\
\hline
PVC & 57.8 & 57.7\\
PTFE & 72.5 & 68.3\\
ESR & 46.7 & 50.5\\
\hline
%\bottomrule
\end{tabular*}
%\end{center}
\end{table}

\subsection{Influence of LS thickness}

Silicone oil was painted on the fiber cross section to improve the coupling between the WLS fiber and the PMT photocathode window. The operation increased the number of collected p.e. by almost 10$\%$. Besides, the thicker the LS, the higher the number of collected p.e. in the detector, making it easier to discriminate between muons and environmental radioactivity. The relationship between the LS thickness and the number of collected p.e. was studied since it can be relevant to optimize future detector designs.

\begin{figure}[htbp]
  \centering
  %\begin{center}
  \includegraphics[width=8cm]{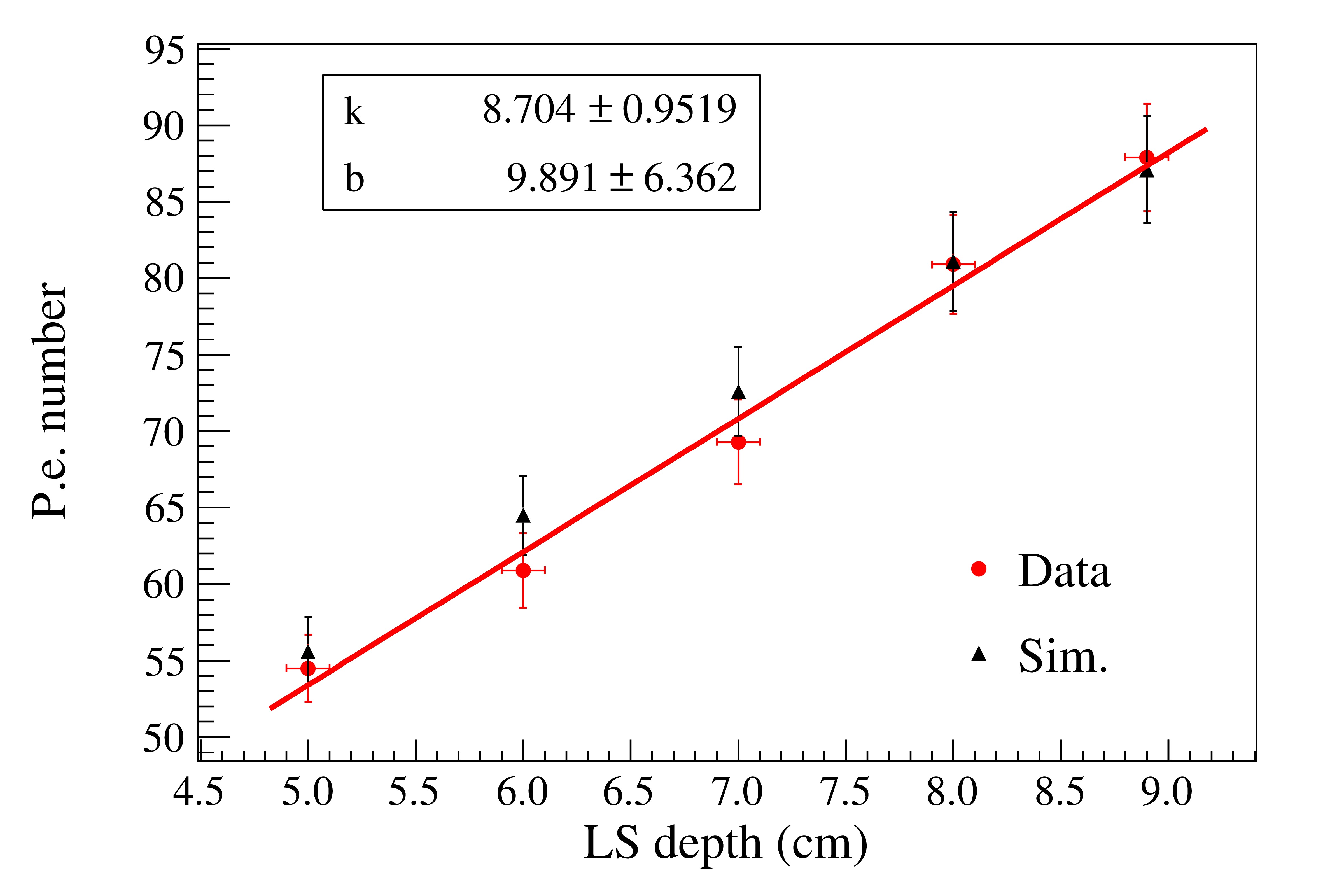}
  \caption{\label{fig:LSdepth} Number of collected p.e versus the LS thickness. Data are compared to the Monte Carlo simulation.}
  %\end{center}
  \end{figure}

We set several depths of LS with PTFE material as reflector to study detector response. Figure \ref{fig:LSdepth} shows the number of collected p.e. as a function of the LS thickness in the LSF detector. The results are compared to the Monte Carlo simulation and show that the results are consistent.

The number of collected p.e. exhibits an almost linear behaviour as a function of the LS thickness. Data have been fitted with a linear function and show an increment of about 8.7 p.e. every increase of 1cm in the LS thickness.

\section{Background discrimination capability}

\begin{figure}[htbp]
  \centering
  %\begin{center}
  \includegraphics[width=8cm]{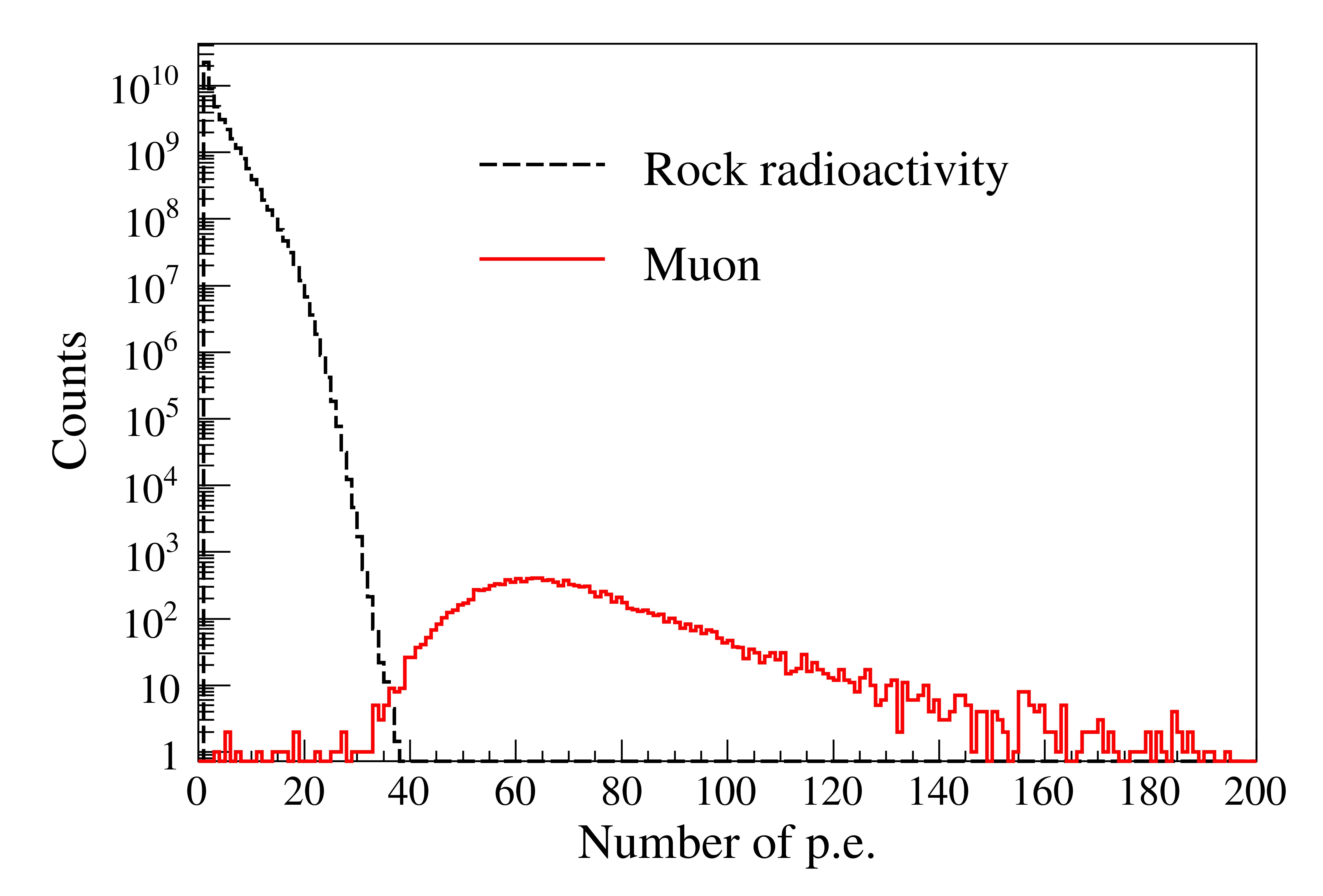}
  \caption{\label{fig:radioactivity} Rock radioactivity and muons response in a detector with 8 cm LS thickness in simulation.}
  %\end{center}
  \end{figure}

If the LSF detector is chosen as an option for a veto detector, cosmic ray muons deposit roughly $\sim$ 2 MeV per centimeter in the LS, while environmental radioactivity is below 3 MeV. Relatively thick liquid scintillator can effectively distinguish muons from radioactivity background which comes from the surrounding rock. Muons and rock radioactivity response in simple LS detector were simulated, as shown in Fig. \ref{fig:radioactivity}. With a 8cm thick LS, the average number of collected p.e. is about 60, while all the radioactivity events from the surrounding rock induce a much lower number of p.e. in the detector. By cutting the events using a 25 p.e. threshold, the muons selection efficiency is larger than 98$\%$. Two layers of detector can be used to suppress further the events rate of the environmental radioactivity. The noise to signal ratio will be smaller than 0.1$\%$ assuming two layers coincidence. The result satisfies the JUNO veto detector requirement~\cite{juno_concept}. For RPC detector, achieving the noise to signal ratio about 0.1$\%$ needs to use at least seven RPC layers~\cite{rpc_layers}. Therefore the LSF detector developed and studied in the the present paper could be chosen as top veto tracker for the JUNO experiment.

\section{Conclusion and discussion}

A Liquid Scintillator with WaveLength Shifting fiber prototype detector has been designed, built, and its performances studied in details. It has been shown that such detector could serve well in identifying cosmic muons traversing the detector and distinguishing them from ambient gamma rays or other such radiation. An average number of 58 p.e. are collected for passing muons, with a detection efficiency greater than 98$\%$. To enhance the reflectivity of the inner walls of the detector, several coating materials have been studied.

By the prototype study, the p.e. collection of ESR with highest reflection is much lower than PVC and PTFE collection. For the results, the ESR reflectivity has been further investigated and dedicated measurements have been performed immersing a ESR film in mineral oil and measuring the reflectivity as a function of the incident angle. The measurement shows that the ESR reflectivity is a sharp function of the incident angle being lower with large incident angle than reflectivity with small incident angle.

Detector with PTFE reflectors shows better performances and 25$\%$ more p.e. collection than with PVC reflectors. PTFE is better suited for the purposes with respect to PVC and ESR. PTFE used as detector container is a good candidate for the future detector designs.

By optimizing the geometry layout structure of the detector, the p.e. collection can be made higher than that of the NO$\nu$A detector~\cite{NOVAproposal} with a superior noise to signal ratio due to the thicker LS in the detector.

Thanks to the outstanding performances of the LSF detector, it has been proposed as an option for the top veto detector of the JUNO experiment. The LSF detector definitely can satisfy JUNO's requirements.

The study can serve as a reference and guideline for future underground low background experiments or accelerator experiments planning to use similar LSF detectors.

\acknowledgments

This work was supported by the Strategic Priority Research Program of the Chinese Academy of Sciences (Grant No.XDA10010300) and the National Natural Science Foundation of China (Grant No.11675203).

% We suggest to always provide author, title and journal data:
% in short all the informations that clearly identify a document.

\end{document}